# Dark Energy and Cosmic Sound: *w(z)* Surveys with the Gemini/Subaru Wide-Field Multi-Object Spectrograph

**Karl Glazebrook (JHU), Daniel Eisenstein (Arizona), Arjun Dey (NOAO), and Bob Nichol (Portsmouth),** for the WFMOS Feasibility Study Dark Energy Team.

**WFMOS Instrument Institutions:** Gemini Observatory, Subaru Observatory

**WFMOS Feasibility Study Team institutions:** Anglo-Australian Observatory, Johns Hopkins University, NOAO, University of Portsmouth, University of Durham, University of Oxford, The Canadian Astronomy Data Centre

## Overview

Precision measurements of the expansion history of the Universe allow us to study directly the evolving influence of dark energy through cosmic time. In recent years, the use of the imprint of primordial baryon acoustic oscillations (BAO) on the late-time clustering of galaxies as a precision 'standard ruler' has come to the forefront as an accurate way to measure the expansion rate of the Universe (Blake & Glazebrook 2003, Hu & Haiman 2003, Seo & Eisenstein 2003, and references in the appended list). Prior to recombination ($z\sim1100$), sound waves propagating in the relativistic photon-baryon fluid imprint a characteristic scale onto the anisotropies of the CMB and the late-time clustering of matter. This scale, about 150 comoving Mpc, is so large as to be essentially immune to low-redshift astrophysics and can be distinctively measured in wide-field galaxy redshift surveys. By measuring the apparent scale of the acoustic signature along and across the line of sight, one can infer the Hubble parameter $H(z)$ at, and the angular diameter distance $D_A(z)$ to the redshift of the target galaxies. Moreover, this distance scale can be connected to $z=1100$ via the angular scale of the CMB acoustic peaks and placed on an absolute distance scale by computing the sound horizon from the matter and baryon density inferred from the CMB. Importantly, the creation and propagation of the acoustic signature is based on simple and well-understood linear cosmological perturbation theory.

In the last 6 years, acoustic oscillations have been observed in the CMB anisotropies at $z\sim1100$ and in the local large-scale galaxy distributions (2DF: Cole et al. 2005; SDSS: Eisenstein et al. 2005). These measurements argue strongly for a low-density, flat Universe dominated by dark energy. The acoustic signature has not yet been measured at intermediate redshifts $0.5<z<3$, where $D_A(z)$ and $H(z)$ provide sensitive and unique probes of dark energy and where the acoustic peaks are easier to measure in the galaxy distribution than at low redshift.

The experiment we propose is to measure the acoustic oscillations at $0.5<z<1.3$ and $2.5<z<3.5$ using a redshift survey of millions of galaxies over 2000 square degrees. This will yield 1–2% precision in the distance-redshift relation and the Hubble parameter, powerfully constraining the characteristics of the dark energy.

This ambitious survey is enabled by a new wide-field (1.5°-diameter field-of-view), highly-multiplexed (4000 fibers) multi-object spectrograph (WFMOS) that has been proposed for the Subaru 8m telescope, and which is envisioned to be funded by a collaboration between the Gemini and Subaru Observatories. WFMOS emerged as a top priority of the U.S. and international Gemini community during the development of a plan for new instrumentation to answer fundamental questions in astronomy. The instrument is briefly described in a 2 page appendix to this document and is fully described in an extensive WFMOS Feasibility Study report (Barden et al. 2005). Its current status is summarized in supporting letters from Gemini and Subaru to the DETF.

## *Baseline Proposal*

| Redshift Range | $R_{lim}$ (AB mag) | Number Density ($h^3$ Mpc$^{-3}$) | Surface Density (deg$^{-2}$) | Total Volume ($h^{-3}$ Gpc$^3$) | Total Area (deg$^2$) | Total Sample Size | Total Survey Time (hrs) |
|---|---|---|---|---|---|---|---|
| $0.5 - 1.3$ | 22.7 | 5 x 10$^{-4}$ | 1000 | 4 | 2000 | 2 x 10$^6$ | 900 |
| $2.3 - 3.3$ | 24.5 | 5 x 10$^{-4}$ | 2000 | 1 | 300 | 6 x 10$^5$ | 800 |

The baseline WFMOS surveys are shown above. The required galaxy number density is set by the need to moderate the effects of shot noise. The $z\sim3$ and $z\sim1$ surveys require the *same* density because the target galaxy populations at both redshifts have similar relative overdensities $\sigma_8\sim0.8$. The survey volume/area is set by the required accuracies in $D_A(z)$ and $H(z)$ needed to facilitate a many-fold improvement over current dark energy constraints. (Our estimates are based on the best available empirical measurements of galaxy number densities.)

The $0.5<z<1.3$ survey will target emission line galaxies. A number density of 5 x 10$^{-4}$ $h^3$ Mpc$^{-3}$ would require 1000 targets per square degree. This number density only requires one to target blue galaxies down to $R$=22.7 even at $z$=1.3. 1400 such pointings would yield 2000 square degrees, which would survey $4h^{-3}$ Gpc$^3$ and 2 million galaxies. For a correlation length of $4h^{-1}$ Mpc , the performance is predicted using Fisher matrix techniques to be 1.0% on $D_A(z)$ and 1.2% on $H(z)$ (1σ). Preselecting spectroscopic targets requires a straight-forward photometric redshift survey, since the Balmer and 4000Å breaks are below 1 μm and hence accessible to optical multi-color imaging surveys. The exposure time for $R$<22.7 emission-line galaxies is calculated to be ~30 min. We budget 40 min per observation to include acquisition overheads. That implies 900 hours of exposure for the survey, or ~90 clear nights. The key redshift diagnostics are at 6000-9000Å, so the observing time need not be particularly dark.

The second survey at $2.3<z<3.3$ will target Lyman-break galaxies. A number density of 5 x 10$^{-4}$ $h^3$ Mpc$^{-3}$, requires 2000 targets per square degree. Sampling a volume of $1.0h^{-3}$ Gpc$^3$ would require a survey of 300 square degrees (600,000 galaxies), or 200 WFMOS pointings. This would yield 1.5% and 1.8% (1σ) accuracies on $D_A(z)$ and $H(z)$ respectively, based on models by Seo & Eisenstein (2003) and including the observed correlation length of $4h^{-1}$ Mpc. The $2.3<z<3.3$ Lyman-break galaxies can be selected using standard photometric techniques (*U*-dropout selection), and existing surveys (e.g., Steidel et al. 1996) suggest a flux limit of $R$<24.5 to achieve the required densities. Exposure times of 4 hours per field will be required, so this survey requires 800 hours, or 80 clear dark nights.

The survey parameters as outlined above are based on our initial optimizations and are not yet set in stone. The *w(z)* constraints outlined below are based on these baseline surveys. We are continuing to explore parameter space, including methods for extending the surveys to $1.3 < z < 2$.

## *Precursor Observations & Calibrations*

This project requires precursor wide-field multi-color imaging surveys to identify targets for the redshift survey. The requirements for the $z \sim 1$ survey are to reach $R \sim 23$ over 2000 deg$^2$ and include two well-measured colors (e.g., *B–R*, *R–I*) in order to allow elimination of $z < 0.5$ foreground galaxies. This depth is easily attainable in about 10 min per band on a 4–meter telescope. The survey should be subdivided around the sky to speed completion, and comprised of regions at least 15$^o$ on a side. The $z \sim 3$ selection requires deeper imaging ($R \sim 25$), but over less area (300 deg$^2$, 10$^o$ on a side). One challenge is the requirement for deep *U*-band observations to allow dropouts to be identified ($\sim 3$-4 hrs per field on a 4m).

A key advantage of the BAO method is that the calibration requirements are not taxing. Essentially this is because the method targets a large preferred scale and hence is highly differential. Galaxy positions need only be accurate to $\sim 1$ arcsec and $\Delta z \sim 0.001$. Redshifts are the only required product of the spectra. The selection of the survey must be reasonably isotropic (or at least tracked in its anisotropies), but imperfections at the few percent level (say, in the relative photometric calibration) can be removed with negligible loss of precision and will not generically mimic the acoustic signature. Similarly, the image quality need only be controlled enough to give reasonably uniform separation of stars from galaxies. It is not necessary that the survey select the identical objects as a function of redshift, as clustering is determined differentially at every redshift and an evolving bias can be measured and removed (and, again, does not mimic the acoustic signature). There is no need for improvement on absolute photometric calibration.

The imaging surveys can be carried out during the design and construction phases of the project. Indeed, several ongoing projects are currently underway which could provide the requisite data. PanSTARRS, Dark Energy Survey (DES), VST ATLAS and KIDS surveys will provide thousands of square degrees of sufficiently deep multi-color optical data. Existing facilities, such as the CFHT Megacam and the NOAO MOSAIC cameras, can map large regions. There are also several near-infrared ground-based options as well that might improve the photometric pre-selection. UKIDSS, NEWFIRM, and VISTA will all be in operation within the next couple of years, and are planning to map large areas. By 2010, it is nearly certain that at least 1000 square degrees of sufficiently deep imaging data will exist.

For preselecting $z \sim 3$ targets, deep *U*- band data are important. This is a challenge because the next generation of major CCD cameras, e.g., PanSTARRS, DES, and perhaps LSST, have poor UV sensitivity. However, the proposed WFMOS redshift survey at $z \sim 3$ is only hundreds of square degrees, not thousands. A concerted effort on existing blue sensitive cameras (e.g., CFHT-MegaCam, NOAO-MOSAIC, VST-OmegaCam or LBT/LBC-blue) should suffice.

## *Systematic Errors*

A particularly attractive characteristic of the BAO method is that it is based on an *absolute* standard ruler, calculated from linear physics as the comoving distance that a sound wave can travel between the time of generation of perturbations and the end of recombination. In the concordance cosmology, this length scale is 150 comoving Mpc. In the correlation function, one finds a single narrow peak at this scale. In the power spectrum, this corresponds to a set of harmonic oscillations, extending to wavenumbers of about $0.3h$ Mpc$^{-1}$. However, most of the distance information is at wavenumbers below $0.2h$ Mpc$^{-1}$. These scales are only barely non-linear today and are in the linear regime at $z>1$.

If the acoustic scale were not known, BAO would still measure the distance scale of the Universe up to an unknown absolute factor. The connection between the acoustic scale as it appears in galaxy clustering and in the CMB must be modeled to connect the distance-redshift relation out to $z=1100$, but this connection is known extremely well from recombination perturbation theory.

The absolute acoustic scale (or sound horizon) depends on the plasma sound speed and the expansion history of the Universe prior to recombination. The first of these is well known from the baryon density measured in the CMB acoustic peaks. The second depends principally on the epoch of matter-radiation equality. In the standard theory, the radiation density is assumed known and one simply measures the matter density from the CMB acoustic peaks. However, modifications from this theory are highly constrained by the CMB anisotropies, and it appears that changes (due to, e.g., neutrino count, baryon density, spectral index) will be either constrained to be negligible or will cancel out from $w(z)$ inferences (Eisenstein & White 2005).

The primary remaining concern is the extent to which non-linear clustering, redshift distortions, and clustering bias affect how galaxies trace the mass on large scales. Indeed, these effects are dominant on small scales, with their strength growing towards lower redshift. All of these effects alter the power spectrum of galaxies on large scales. However, none of them create a preferred scale that might mimic, confuse, or bias the acoustic scale. Broad-band tilts of the power spectrum may shift the apparent scale, but in ways that are trivial to restore.

However, while these non-linearities do not shift the acoustic scale, they may erase or obscure it, causing the signal-to-noise estimates to be optimistic. Estimates of this effect have been included in all our Fisher analyses to date and are now being validated by various cosmological simulations. These simulations show that the non-linear clustering and redshift distortions erase the higher harmonics at low redshift, roughly down to $k=0.15h$ Mpc$^{-1}$ at $z=0.5$. However, at $z>1$, the acoustic signal is well preserved.

More generally, galaxies and dark matter halos exhibit a clustering bias relative to the full dark matter distribution. This bias is expected to be scale-independent on large scales (Coles 1993; Scherrer & Weinberg 1998), preserving the acoustic signal, but is generically scale-dependent on small-scales. N-body simulations, whose only physics is gravity, have confirmed that the power spectrum of dark matter halos preserves the acoustic oscillation signature. Seo & Eisenstein (2005) show that simple halo bias models preserve the large-scale acoustic signature to better than the required accuracy level. More complex schemes using both semi-analytic and hydrodynamical models of galaxy formations give similar results (Angulo et al. 2005; Springel et al. 2005; Figure 2). This is not a surprise; it arises simply because galaxy formation does not create a preferred scale anywhere above the halo size scale.

The consensus from numerous authors (and our Feasibility Study) is that there are no significant systematic issues for WFMOS-scale surveys for acoustic oscillations. Provided the volume/number can be achieved, the $w(z)$ constraints are achievable.

While the acoustic oscillation method is theoretically and observationally clean, it does require very large volumes to detect the peaks. The statistical error on the clustering measurement is limited simply by the survey volume and the number of galaxies. There is a critical number density of galaxies, which scales inversely with the power spectrum amplitude (i.e., as $\sigma_8^{-2}$), above which the measurement is essentially volume-limited (and *not* number-limited). This drives the experiment to very wide fields. The required number density is well below that of $L^*$ galaxies; the targets are always the bright galaxies at a given redshift.

## Quantifying Dark Energy Results

Using the baseline WFMOS survey, we use a Fisher matrix analysis to predict the constraints on the angular diameter distance and Hubble parameter. We find 1.0% (1.2%) on $D_A$ ($H$) at $z\sim1$ and 1.5% (1.8%) at $z\sim3$ (all 1-$\sigma$). It is important to note that methods that rely on the distance-redshift relation alone must differentiate $D_A(z)$ to get to $H(z)$ and thereby to the dark energy equation of state. The direct constraints on $H(z)$ resulting from this BAO survey therefore compare better than their number would indicate; for example, the 1.2% measure of $H(z=1)$ corresponds to a 0.7% measure of $D_A(z=1.7)/D_A(z=0.8)$, i.e., 1.5% in a flux ratio. These accuracies on $H(z)$, $D_A(z)$ have been verified by an independent code based on Monte-Carlo techniques and are very robust (e.g., Table 1 of Glazebrook and Blake 2005).

We then combine these measurements with Planck-equivalent CMB forecasts and the SDSS LRG sample to predict constraints on $w(z)$. Perturbing around $w=-1$ in a flat cosmology, we would measure $w$ to within 8% (for a constant $w$) and $w_1$ (= $dw/dz$) to 25% (1$\sigma$). This is shown in Figure 3. Including a ground-based supernovae program that measures 1% distances per $\Delta z=0.1$ bin out to $z=1$ reduces the errors to 5% and 20%. If one extends this SNe sample out to $z=1.7$, as a Joint Dark Energy Mission (JDEM) might do, then one finds 3.5% and 22% performance ignoring WFMOS and SDSS. In other words, the WFMOS surveys combined with ground-based supernovae studies do well compared to a space-based supernova program.

These are conservative estimates regarding the impact of acoustic oscillation surveys on dark energy because the $w=-1$ fiducial model is a choice that favors low-redshift methods. Moreover, the acoustic oscillation method offers the important opportunity to probe $H(z)$ at $z>2$, including the comparison of distances out to $z=1000$. Having been surprised by the mere existence of dark energy, one should measure dark energy at high redshift rather than simply assume it to be small.

Finally, the acoustic constraints are extremely powerful for exploring the possibility of small non-zero curvatures that might otherwise confuse the interpretation of $w(z)$. This advantage arises because of the ability to connect the low-redshift distance scale out to $z=1000$. This is shown in Figure 4 as well as in Figure 2 from Bernstein (2005). It would be a severe error to mistake a cosmological constant with a small spatial curvature for a $w=-0.95$ flat cosmology.

## Project Risks

i. The project requires pre-imaging. Since many such surveys are being proposed and started now, which are also motivated by dark energy (via lensing), the risk of not having imaging data available is low.

ii. The $w(z)$ constraints assume Planck CMB accuracy that may not come to pass. However WMAP (4 years data) is adequate; having to rely on WMAP first-year data alone would degrade the accuracy.

iii. The WFMOS Feasibility Study identified the wide-field corrector as being the highest risk component. The corrector front element has a clear aperture of 1.2 to 1.4 meters, comparable in size to the correctors envisioned for other dark energy experiments. The feasibility study for the corrector is still under way.

iv. The instrument is expensive ($45M) for a ground-based instrument. To mitigate this risk, Gemini Observatory is including a 30% contingency.

v. The instrument construction is contingent on partnership and joint funding with Subaru.

## *Project Strengths*

i. The acoustic oscillation scale is a fossil from the early Universe and does not evolve at late times. Its absolute scale can be calculated precisely from high-redshift linear perturbation theory and CMB measurements, although the method is robust even to interesting changes in the early Universe.

ii. The acoustic scale is robust against astrophysical complications. The minor effects that do exist can be quantified by cosmological simulations.

iii. The measurement of the acoustic scale is observationally straight-forward. We expect to be limited only by sample variance.

iv. A standard ruler approach is insensitive to dust or other more exotic photon-destroying processes (e.g., axion coupling) that could affect standard candles. Similarly, it is insensitive to magnification by gravitational lensing.

v. An internal cross-check is built-in to the method, as $D_A(z)$ must be the appropriate integral of $H(z)$.

vi. The distance-redshift relation is on the same scale as the CMB measurement of the distance to $z=1100$. This provides sensitivity for any dark energy effects at $3<z<1100$ and permits the isolation of non-zero spatial curvature from inferences of $w(z)$.

vii. Acoustic oscillations can in principle be traced to any redshift where galaxies can be seen. The WFMOS baseline experiment would probe the distance scale and Hubble parameter at $z=3$, unlike any other ground-based method we know of.

viii. WFMOS surveys would also permit measurement of the Alcock-Paczynski effect, which would add additional constraints on dark energy provided that the redshift distortions can be separated from the cosmological distortion.

ix. WFMOS will be the premier wide-field spectrograph in the world and as such is well-positioned to optimize its survey strategy to take account of developments in dark energy over the next 5 years. WFMOS surveys will also be the best calibrator of photometric redshifts to support cluster and weak lensing dark energy surveys.

x. The spectroscopic datasets resulting from the $w(z)$ surveys will enable a host of extragalactic science (e.g., studies of galaxy formation and evolution, chemical evolution, clustering, etc.) and have an enduring legacy.

xi. WFMOS is an international project, including partners from US, UK, Canada, Australia,

and Japan. This provides an exceptional array of talent and resources for undertaking the preimaging surveys, scientific planning and instrument design and construction.

## Technology R&D

No serious technical or R&D risks were identified in the WFMOS Feasibility Study (other than those mentioned above) and we are now proceeding to the concept design phase.

## Relationship to JDEM and LSST

Neither JDEM or LSST are critical to WFMOS achieving its goals, although in principle they could provide the requisite preimaging. The required imaging is shallow compared to LSST, so even the earliest LSST imaging could provide the input catalogs. WFMOS is likely to precede JDEM (and perhaps LSST), and could provide the spectroscopic follow-up capability in support of LSST and JDEM surveys. Since WFMOS would be the precursor to any JDEM acoustic oscillation experiment, and since WFMOS surveys would approach JDEM levels of precision in the distance–redshift measurement, any discovery could inform the JDEM mission strategy.

## Relation to other Acoustic Oscillation Experiments

A number of other near-term acoustic oscillation projects are being considered, but none currently planned will be able to approach the capabilities of the WFMOS survey. The survey speed of the WFMOS instrument (field-of-view, multiplexing, and instrument sensitivity) is 10-30 times larger than any existing, funded or planned facility.

It is indeed possible to pursue acoustic oscillations with photometric redshift surveys. However, these surveys require 10-20 times more sky coverage to reach equivalent specifications, making them essentially full sky. These surveys cannot recover $H(z)$ directly, which is a significant loss, particularly at high redshift. Finally, photometric redshift surveys are more susceptible to systematic errors in their calibration. Hence, we have concluded that spectroscopic surveys, targeting multiple redshift ranges, are preferred.

## Facilities Needed

WFMOS requires an 8-m class telescope in order to be able to measure redshifts sufficiently fast to deliver results in a few years. We have working designs for both Gemini and Subaru, of which Subaru is preferred as it is better matched to prime focus instrumentation. The concept could be readily generalized to other large telescopes.

To plan and conduct a large, cohesive redshift survey, and analyze and distribute the results requires a major undertaking by a dedicated science team. Although the funding for the instrument design and construction is likely to materialize, the funding path for the science operations has yet to be identified. Given the need to organize the precursor imaging and plan the spectroscopic surveys, one cannot wait until the instrument is commissioned to begin.

We note that the data analysis requirements are modest (a few terabytes of data) compared to the next generation of imaging surveys.

## Experiment timeline

2005-2006  Concept Design Study

2007-2009  Preliminary and Final Design

2010-2012  WFMOS Build and Commissioning

2013-2016  Dark Energy Science Survey (Early results from first 500 deg$^2$ by end 2013)

## References

This list of references is incomplete. We have focused on documenting the recent work on the acoustic oscillation method and included other references only when required to document important assertions.

The instrument concept and BAO key science was first discussed in the KAOS Purple Book (http://www.noao.edu/kaos/KAOS_Final.pdf). This formed the basis of the Feasibility Study commissioned by Gemini (Barden et al., 2005).

Some papers in the history of predictions of the acoustic phenomenon: Peebles, P.J.E. & Yu, J.T. 1970, ApJ, 162, 815; Sunyaev, R.A., & Zel'dovich, Ya.B., 1970, ApSS, 7, 3; Bond, J.R. & Efstathiou, G. 1984, ApJ, 285, L45; Holtzmann, J.A. 1989, ApJ Supp, 71,1; Hu, W., & Sugiyama, N. 1996, ApJ, 471, 542; Bashinsky, S., & Bertschinger, E., 2001, PRL, 87, 1301

There have been many papers on forecasting the constraints available from the low-redshift acoustic-scale measurement: Eisenstein, D. J., Hu, W., & Tegmark, M., 1998, ApJ, 504, L57; Eisenstein, D. J., Hu, W., & Tegmark, M., 1999, ApJ, 518, 2; Eisenstein, D.J., 2003, in Wide-field Multi-Object Spectroscopy, ASP Conference Series, ed. A. Dey; Blake, C., & Glazebrook, K., 2003, ApJ, 594, 665; Hu, W. & Haiman, Z., 2003, PRD, 68, 3004; Linder, E.V., 2003, PRD, 68, 3504; Seo, H., & Eisenstein, D.J., 2003, ApJ, 598, 720; Amendola, L., Quercellini, C., & Giallongo, E., 2005, MNRAS, 357, 429, astro-ph/0404599; Blake, C., Abdalla, F., Bridle, S., & Rawlings, S., 2004, New Astron. Rev., 48, 1063; Matsubara, T., 2004, ApJ, 615, 573; Glazebrook, K., & Blake, C., 2005, ApJ, in press, astro-ph/0505608

Cole, S., et al., 2005, MNRAS, submitted, astro-ph/0501174; Eisenstein, D.J., et al., 2005, ApJ, submitted, astro-ph/0501171

Papers on using photometric redshifts for distance-scale measurements: Cooray, A., Hu, W., Huterer, D., & Joffre, M., 2001, ApJ, 557, L7; Seo, H., & Eisenstein, D.J., 2003, ApJ, 598, 720; Blake, C., & Bridle, S., 2004, MNRAS, submitted, astro-ph/0411713; Dolney, D., Jain, B., & Takada, M., 2004, astro-ph/0409445

Eisenstein, D.J., & White, M., 2004, PRD, 70, 103523

Bernstein, G., 2005, ApJ, submitted, astro-ph/0503276

Papers relevant to the survival of the acoustic features in cosmological simulations: Meiksin, A., White, M., & Peacock, J.A., 1999, MNRAS, 304, 851; Angulo et al., 2005, MNRAS,

# Figures

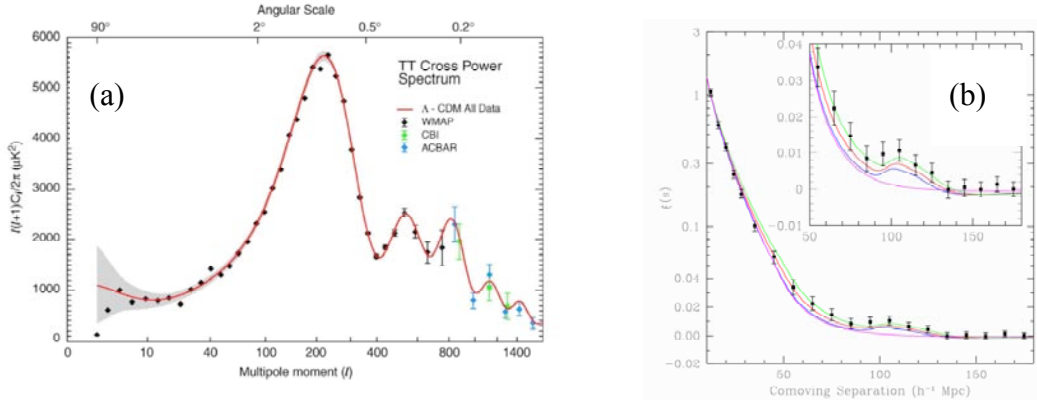

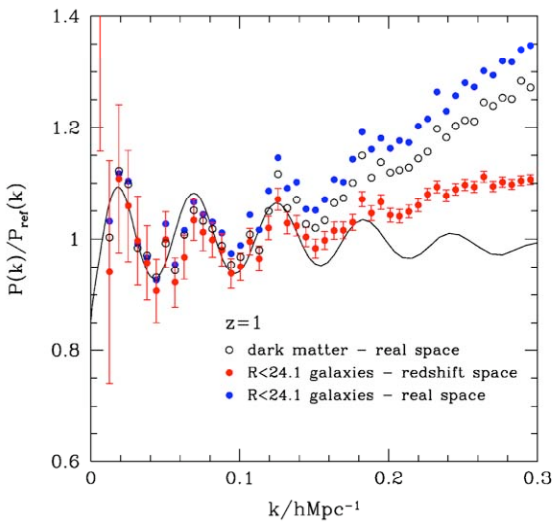

**Figure 1**. Baryon acoustic oscillations in (a) the CMB temperature fluctuation power spectrum (Bennett et al. 2003), (b) the SDSS Luminous Red Galaxy Survey (Eisenstein et al. 2005) and (c) in a simulated WFMOS survey of 600 deg$^2$ at $0.5<z<1.3$ and 2 million galaxy redshifts.

**Figure 2**. The power spectrum of dark matter and galaxies (points) compared to linear theory (solid line) at $z=1$ from the Millennium simulation (Springel et al. 2005) with galaxies inserted using the semi-analytic approach of Baugh et al 2005. On large scales the linear peaks are preserved (see WFMOS feasibility report for more details + other similar tests).

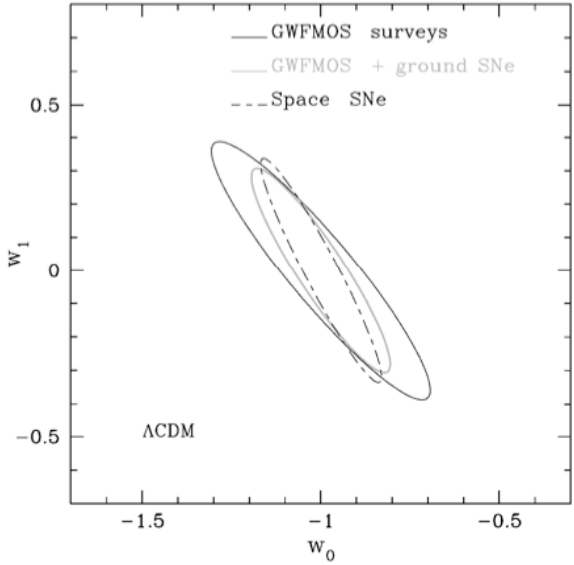

**Figure 3**. The performance on a model space of $w(z)=w_0 + w_1z$ from the baseline WFMOS surveys of 2000 square degrees at $z\sim1$ and 300 square degrees at $z\sim3$, including final SDSS and Planck constraints. We then combine the GWFMOS constraints with aggressive ground-based supernova performance: 1% in distance for 9 $\Delta z=0.1$ bins up to $z=1$. Were $w_1$ held at 0, then the error on $w$ would be 5%. However, $w_1$ opens a difficult degeneracy for all methods. We compare the result to a mock JDEM satellite mission, with 16 1% distance bins up to $z=1.7$. This gives similar performance

**Figure 4**. The constraints in the space of constant w and non-zero curvature from various measurements. The green line shows the 3% measurement of the distance to $z=0.35$ expected from the SDSS. The solid and dashed black lines show the 1% and 1.2% constraints on $D_A(1)$ and $H(1)$ from the WFMOS survey. The blue line shows a 1% measurement of the ratio of the distance to $z=0.8$ to that to $z=0.05$; this is equivalent to a 2% flux ratio measurement for supernovae out to $z=0.8$. One sees that acoustic oscillations are highly complementary with $z<1$ supernovae in this parameter space. One also sees that assuming flatness (as opposed to testing it) benefits the SNe constraint on w considerably more than the acoustic oscillation constraint.

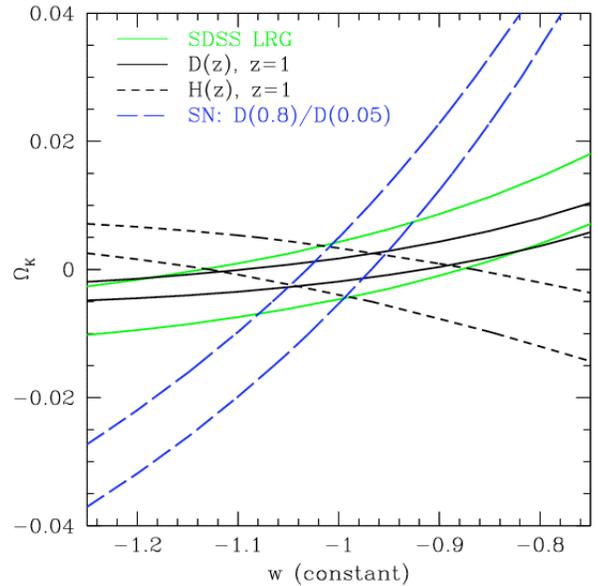

# Appendix: WFMOS Instrument Brief Description

## *Primary Description*

The Wide-Field Multi-Fiber Multi-Object Spectrograph (WFMOS) is an optical spectrograph for very high multiplex observing over a very wide field of view on an 8 m telescope. The baseline instrument is described as follows:

- Implementation at the prime focus of the Subaru telescope

- A 1.5 degree diameter field of view

- Wavelength passband of 0.39-1.0 microns

- ~4500 fiber apertures positioned by 'Echidna' piezo-actuator mechanism. (Currently being used in FMOS instrument for Subaru).

    - ~3000 dedicated to a bank of low-dispersion spectrographs

    - ~1500 dedicated to a bank of high-dispersion (R=40000) spectrographs (not applicable to the dark energy science application)

- ~1 arcsecond apertures

- Set of ~10 low-dispersion spectrographs based upon same design as SDSS spectrographs providing spectral resolutions of R=1800 in the blue and R=3500 in the red with good throughput.

- Nod & Shuffle observing mode for optimal sky subtraction

- Reduced data archived in Gemini archive (available for ancillary science)

Figure A1 shows how this instrument would fit onto the Subaru telescope with the wide field corrector feeding an Echidna fiber positioner (Figure A2) and the fibers feeding the bank of spectrographs (Figure A3) located off to the side of the telescope building. Figure A4 shows the instrument efficiency, which surpasses the SDSS fiber spectrographs (similar optics but better CCDs and grisms).

The WFMOS feasibility report (Barden et al., 2005) provides a comprehensive description of the science case and the instrument concept.

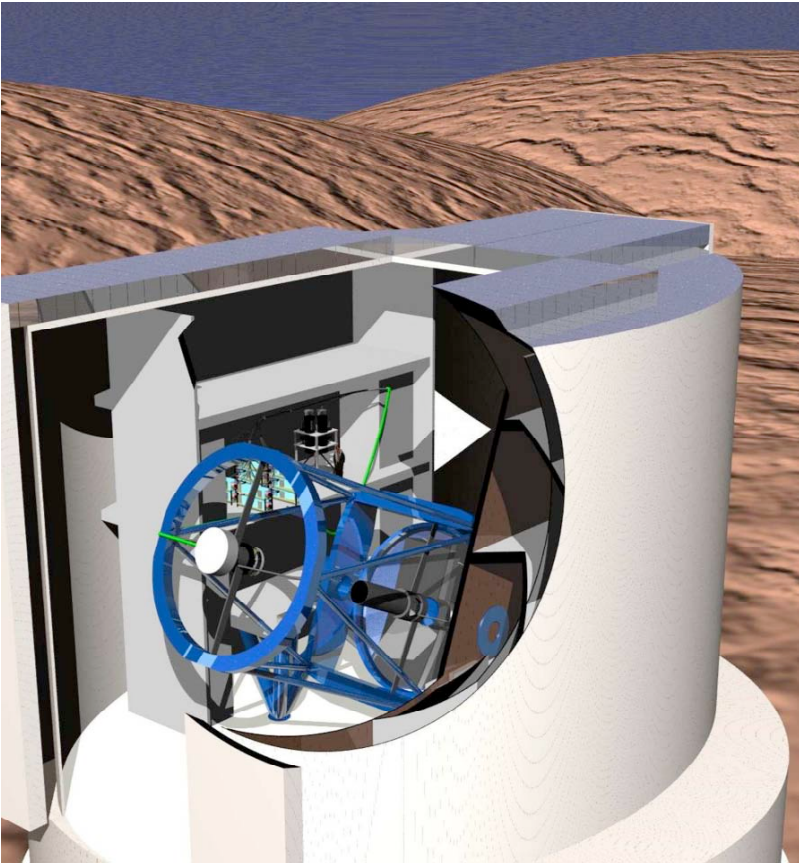

**Figure A1 (left).** CAD rendering of WFMOS on the Subaru telescope. The 4000-fiber positioner is at prime focus, the green line shows the fiber route down to 10 low-resolution (and 2 high) spectrographs on the 'Subaru Great Wall)'

**Figure A2 (below).** The FMOS-Echidna fiber positioner consists of a dense array of some 400 spines mounted on ball-pivots in piezoelectric actuators. Identical linear modules, each carrying a double row of actuators, make up the square array. Fibers are carried through to the tips of the spines to reach the telescope focal surface. The actuators can be operated simultaneously to position all fibers to 10μm accuracy.

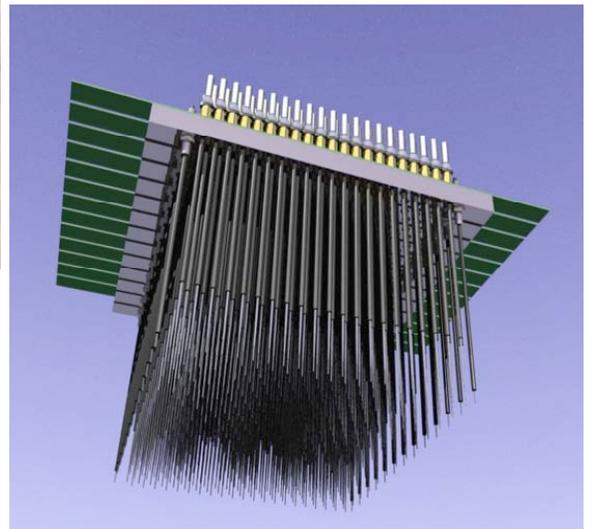

**Figure A3.** Rendering of one of the 10 WFMOS low resolution spectrographs. The design has been scaled from the operational SDSS optomechanical design to meet WFMOS requirements.

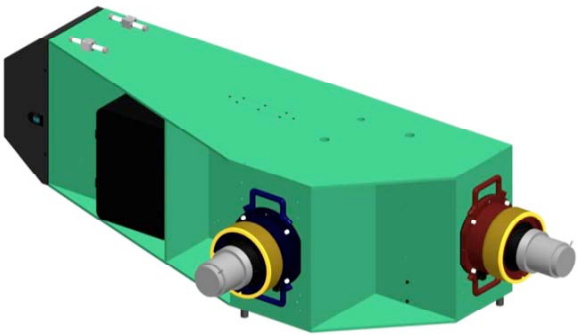

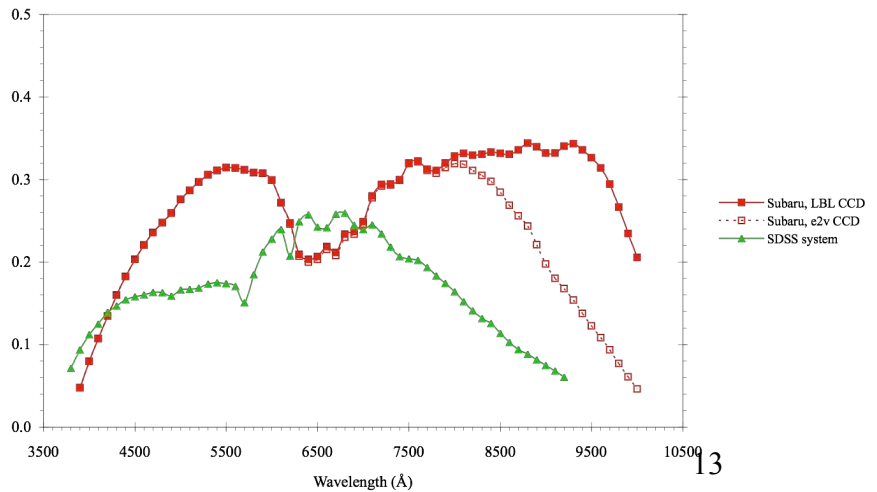

WFMOS Low Dispersion Spectrograph System Efficiency

- Subaru, LBL CCD
- Subaru, e2v CCD
- SDSS system

**Figure A4.** The total system efficiency for WFMOS (including telescope and atmosphere but not including slit losses). Although this prediction is very high, the design and predictions are grounded on the existing SDSS spectrographs.